\documentclass[11pt]{article}

\newif\ifblindversion
\blindversionfalse

\usepackage{relsize}
\usepackage{caption}
\usepackage{float}
\usepackage{subcaption}

\usepackage{amsmath,amsfonts,amsthm}
\begingroup
\makeatletter
\@for\theoremstyle:=definition,remark,plain\do{%
  \expandafter\g@addto@macro\csname th@\theoremstyle\endcsname{%
    \addtolength\thm@preskip\parskip
  }%
}
\endgroup
\usepackage{siunitx}
\theoremstyle{plain}
\newtheorem{proposition}{Proposition}
\newtheorem{remark}{Remark}
\usepackage[sc]{mathpazo}
\usepackage[T1]{fontenc}
\usepackage{xcolor}
\usepackage{graphicx}
\usepackage{epstopdf}
\usepackage[top=2.5cm, bottom=2.5cm, left=2.5cm, right=2.5cm]{geometry}
\usepackage{parskip}
\usepackage{setspace}
\usepackage[unicode=true,bookmarks=false,pdfstartview={FitH},colorlinks=true,citecolor=blue,linkcolor=blue,urlcolor=blue]{hyperref}
\usepackage{mathtools}
\mathtoolsset{showonlyrefs}
\usepackage{enumitem}
\setlist[enumerate]{leftmargin=*}
\setlist[itemize]{leftmargin=*}
\usepackage{natbib}
\usepackage[runin]{abstract}
\abslabeldelim{.}
\usepackage{bm}
\usepackage{xspace}

\makeatletter
\newcommand{\ud}{\mathop{}\!{\operator@font{d}}}
\newcommand{\iid}{\stackrel {\operator@font{iid}}{\sim}}
\newcommand{\ind}{\stackrel {\operator@font{ind}}{\sim}}
\makeatother
\renewcommand{\a}{\alpha}
\newcommand{\g}{\gamma}
\newcommand{\abs}[1]{\left \vert #1 \right \vert}
\newcommand{\fder}[3][]{\frac{\mathrm{d}^{#1}#3}{\mathrm{d} {#2}^{#1}}}
\newcommand{\intii}{\int_{-\infty}^\infty}
\newcommand{\sumi}[3][i]{\sum_{#1 = #2}^{#3}}
\DeclareMathOperator*{\argmax}{arg\,max}
\DeclareMathOperator*{\argmin}{arg\,min}
\def\bsqrt{\mathpalette\DHLhksqrt}
\def\DHLhksqrt#1#2{%
\setbox0=\hbox{$#1\sqrt{#2\,}$}\dimen0=\ht0
\advance\dimen0-0.2\ht0
\setbox2=\hbox{\vrule height\ht0 depth -\dimen0}%
{\box0\lower0.4pt\box2}}
\DeclareMathOperator{\var}{var}
\DeclareMathOperator{\cov}{cov}

\newcommand{\rv}[3][1]{#2_{#1},\ldots,#2_{#3}}
\renewcommand{\|}{\,|\,}
\newcommand{\N}{\mathcal N}

\renewcommand{\S}{\mathcal{S}}
\newcommand{\F}{\mathcal{F}}
\newcommand{\f}{f}

\newcommand{\kbt}{k_BT}
\newcommand{\X}{{\bm X}}
\newcommand{\Y}{{\bm Y}}
\newcommand{\dt}{{\Delta t}}
\newcommand{\fs}{{\f_s}}
\renewcommand{\t}{{\bm \theta}}

\newcommand{\M}{{\mathcal M}}
\newcommand{\Aw}{A_\textrm{w}}
\newcommand{\Rw}{R_\textrm{w}}
\newcommand{\Af}{A_\textrm{f}}
\renewcommand{\G}{\mathcal G}
\newcommand{\ee}{{\bm \eta}}
\newcommand{\obj}{\mathcal Q}
\newcommand{\E}{E}
\newcommand{\spp}[1]{^{(#1)}}
\newcommand{\baseline}{{\textnormal{base}}}
\newcommand{\noise}{{\textnormal{noise}}}
\newcommand{\denoise}{{\textnormal{corr}}}
\newcommand{\LP}{{\textnormal{LP}}\xspace}
\newcommand{\NLS}{\textnormal{NLS}\xspace}
\newcommand{\MLE}{\textnormal{MLE}\xspace}
\newcommand{\grantnumber}{RGPIN-2014-04225}

\title{Robust and Efficient Parametric Spectral Estimation in \\ Atomic Force Microscopy}
\date{June 1, 2017}
\ifblindversion
\author{}
\else
\renewcommand\footnotemark{}
\author{\begin{minipage}{.8\textwidth}
\begin{center}
Bryan Yates$^\dagger$  \hfill Aleksander Labuda$^\ddagger$ \hfill Martin Lysy$^{\dagger\star}$
\end{center}
\end{minipage}
\thanks{$^\dagger$Department of Statistics and Actuarial Science, University of Waterloo, Waterloo, ON.}\thanks{$^\ddagger$Asylum Research, an Oxford Instruments Company, Santa Barbara, CA.}\thanks{$^\star$mlysy@uwaterloo.ca. Research supported by NSERC grant \grantnumber.}}
\fi

\begin{document}

\maketitle

\begin{abstract}
  An atomic force microscope (AFM) is capable of producing ultra-high resolution measurements of nanoscopic objects and forces.  It is an indispensable tool for various scientific disciplines such as molecular engineering, solid-state physics, and cell biology.  Prior to a given experiment, the AFM must be calibrated by fitting a spectral density model to baseline recordings.  However, since AFM experiments typically collect large amounts of data, parameter estimation by maximum likelihood can be prohibitively expensive.  Thus, practitioners routinely employ a much faster least-squares estimation method, at the cost of substantially reduced statistical efficiency.  Additionally, AFM data is often contaminated by periodic electronic noise, to which parameter estimates are highly sensitive.  This article proposes a two-stage estimator to address these issues.  Preliminary parameter estimates are first obtained by a variance-stabilizing procedure, by which the simplicity of least-squares combines with the efficiency of maximum likelihood.   A test for spectral periodicities then eliminates high-impact outliers, considerably and robustly protecting the second-stage estimator from the effects of electronic noise.   Simulation and experimental results indicate that a two- to ten-fold reduction in mean squared error can be expected by applying our methodology.

\end{abstract}


\section{Introduction}\label{sec:intro}

An atomic force microscope (AFM) is a scientific instrument producing high-frequency ultra-precise displacement readings of a minuscule ($\SI{\sim 100}{\micro\meter}$-long) pliable beam referred to as a cantilever.  The cantilever bends in response to various forces exerted by its surrounding environment, the recording of which has been immensely useful for the study of e.g.,
the composition of polymers and other chemical compounds~\citep{sugimoto.etal07,garcia.etal07}, interatomic and intramolecular forces~\citep{radmacher97,hoffmann.etal01}, pathogen-drug interactions~\citep{alsteens.etal08}, cell adhesion~\citep{evans.calderwood07}, and the dynamics of protein folding~\citep{yu.etal17}.

\begin{figure}[!htb]
  \centering
  {\includegraphics[width=\textwidth]{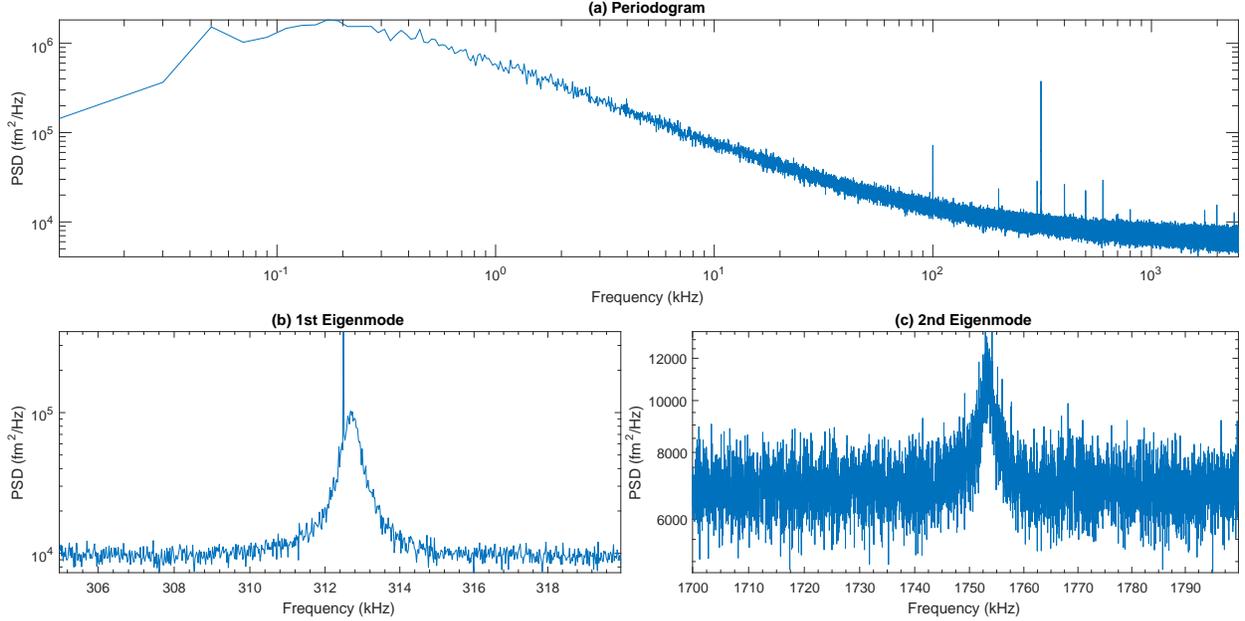}	\phantomsubcaption\label{fig:psdfull}\phantomsubcaption\label{fig:psdeig1}\phantomsubcaption\label{fig:psdeig2}}
  \caption{(a) Periodogram for an AC160 (Olympus) cantilever recorded for \SI{5}{\second} at \SI{5}{\mega\hertz} ($N = \num{25e6}$ observations).  The data have been averaged by bins of size $B = 100$ to enhance visibility.  (b-c) Magnified view of first and second eigenmodes.}
  \label{real_data_psd}
\end{figure}

In a typical AFM experiment, the cantilever's bending response is measured in opposition to its spring-like restoring force, which requires proper calibration of the cantilever stiffness in order to convert measured displacement readings into force~\citep{cleveland.etal93,burnham.etal02,clarke.etal06,sader.etal11}.  This calibration is accomplished by fitting various parametric models to a baseline spectral density recording, i.e., to a cantilever driven by thermal noise alone.  A representative baseline spectrum calculated from experimental data is displayed in Figure~\ref{real_data_psd}.
This data serves to illustrate two outstanding challenges in AFM parametric spectral density estimation. First, the experiments produce massive amounts of data, for which maximum likelihood estimation can be prohibitively expensive.  A much faster least-squares method is routinely employed in practice~\citep{norrelykke.flyvbjerg10} -- at the cost of substantially reduced statistical efficiency. Second, parametric estimates by either method are severely affected by electronic noise, due to periodic fluctuations in the AFM's circuitry.  Such noise is evidenced by the presence of sharp peaks (i.e., vertical lines) in the baseline spectrum of Figure~\ref{real_data_psd}.

In this article, we propose a two-stage estimator addressing both of these issues.  A preliminary estimator first applies a variance-stabilizing transformation which renders the least-squares estimator virtually as efficient as the MLE.  After the preliminary fit, an automated denoising procedure, based on a well-known statistical test for hidden periodicities, robustly protects the second-stage estimator from most of the effects of electronic noise.
Extensive simulations and experimental results indicate that a two- to ten-fold reduction in mean squared error can be expected by applying our methodology.

The remainder of the paper is organized as follows. Section~\ref{sec:psd} provides an overview of parametric spectral density estimation in the AFM context. Section~\ref{sec:est} describes our proposed two-step estimator.  Section~\ref{sec:sim} presents a detailed simulation study comparing our proposal to existing methods.   Section~\ref{sec:data} applies the methodology to calibration of a real AFM cantilever and we close in Section~\ref{sec:disc} with a discussion of future work.

\section{Parametric Spectral Density Estimation in AFM}\label{sec:psd}

Let $X_t = X(t)$ denote a continuous stationary Gaussian stochastic process with mean $\E[X_t] = 0$ and autocorrelation $\g(t) = \cov(X_s, X_{s+t})$.  The power spectral density (PSD) of $X_t$ is then defined as the Fourier transform of its autocorrelation,
\begin{equation}\label{eq:psd}
\S(\f) = \intii e^{-2\pi i t \f} \g(t) \ud t = \F\{\g(t)\}.
\end{equation}
Spectral densities can be used to express the solutions of various differential equations and are thus commonly employed in many areas of physics.  A particularly important example for AFM applications is that of a \emph{simple harmonic oscillator} (SHO). This model for the thermally-driven tip position $X_t$ of the AFM cantilever is
\begin{equation}\label{eq:sho}
m \ddot X_t = - k X_t - \varsigma \dot X_t + F_t,
\end{equation}
where $\dot X_t$ and $\ddot X_t$ are velocity and acceleration, $m$ is the tip mass, $k$ is the cantilever stiffness, $\varsigma$ is the viscous damping from the surrounding medium (e.g., air, water), and $F_t$ is the thermal force which drives the cantilever motion.  It is a stationary white noise process with $\E[F_t] = 0$ and $\cov(F_s, F_{s+t}) = 2\kbt\varsigma \cdot \delta(t)$, where $T$ is temperature and $k_B$ is Boltzmann's constant. While the autocorrelation of $X_t$ has no simple form, a straightforward calculation in the Fourier domain obtains the spectral density
\begin{equation}\label{eq:shopsd}
\S(\f) = \frac{\kbt/(k\cdot \pi\f_0Q)}{\big[(\f/\f_0)^2-1\big]^2 + \big[\f/(\f_0Q)\big]^2},
\end{equation}
where $\f_0 = \bsqrt{k/m}/(2\pi)$ is the cantilever's resonance frequency and $Q = \bsqrt{km}/\varsigma$ is its ``quality factor'', which measures the width of the PSD amplitude peak around $\f_0$ (see Figure~\ref{fig:sim_psd}).

\begin{remark}
The presentation above glosses over several technical details, e.g., non-integrability of many well-defined autocorrelations in~\eqref{eq:psd} (such as that of the SHO), and limited interpretability of the white noise process $F_t$ in~\eqref{eq:sho} as a function of $t$.  For a rigorous treatment of these issues see~\cite{ito54}.
\end{remark}

\subsection{Parametric Inference}

In a parametric setting, the PSD is expressed as $\S(\f, \t)$ and the goal is to estimate the unknown parameters $\t$ from discrete observations $\X = (\rv [0] X {N-1})$ recorded at sampling frequency $\f_s$, such that $X_n = X(n \cdot \dt)$ and $\f_s = 1/\dt$.  Ideally one would work directly with the loglikelihood in the time domain, $\ell(\t \| \X)$.  However, this approach is inviable in practice since it (i) requires Fourier inversion of the PSD to obtain the variance of $\X$, and (ii) scales quadratically in the number of observations \citep[e.g.,][Proposition 8.2.1]{brockwell.davis13}.  Instead, parametric inference can be considerably simplified by making use of the following result.
\begin{proposition}\label{prop:expo}
Let $N = 2K+1$ and define the finite Fourier transform $\tilde {\X} = (\tilde X_{-K}, \ldots, \tilde X_0, \ldots, \tilde X_K)$ of $\X$ as
\[
\tilde X_k = \sumi [n] 0 {N-1} e^{-2\pi i  k n/N} X_n.
\]
For each $\tilde X_k$, let $\f_k = \tfrac k N \f_s$ denote the corresponding frequency.  Then if $\S(\f)$ is the PSD of $X_t$, under suitable conditions on $\S(\f)$, and as $N \to \infty$ and $\dt \to 0$, we have
\[
\tfrac 1 N \abs{\tilde X_k}^2 \ind \f_s \cdot \S(\f_k) \times \textnormal{Expo}(1), \qquad 0 \le k < K.
\]
\end{proposition}
Proposition~\ref{prop:expo} leads to the so-called Whittle loglikelihood function~\citep{whittle57}
\begin{equation}\label{eq:whittle}
\ell_W(\t \| \Y) = -\sum_{k=1}^K \big(Y_k/\S_k(\t) + \log \S_k(\t)\big),
\end{equation}
where $Y_k = \tfrac 1 N \abs{\tilde X_k}^2$ and $\S_k(\t) = \f_s \cdot \S(\f_k, \t)$.  Since the periodogram $\Y = (\rv Y K)$ can be computed in $\mathcal O(N \log N)$ time using the Fast Fourier Transform, maximization of the Whittle loglikelihood~\eqref{eq:whittle} is considerably easier than of the original likelihood $\ell(\t \| \X)$.  Conditions for the convergence of the Whittle MLE $\hat \t_W = \argmax_\t \ell_W(\t \| \Y)$ to the true MLE $\hat \t = \argmax_\t \ell(\t \| \X)$ have been established by~\cite{fox.taqqu86,dahlhaus89}.  Since the true MLE is typically unavailable, we shall refer to Whittle's $\hat \t_W$ simply as the MLE in the developments to follow.

\subsection{Periodogram Binning}\label{sec:bin}

Despite its computational advantages relative to exact maximum likelihood, obtaining $\hat \t_W$ often remains a practical challenge, due to the enormous size of typical AFM datasets and the difficult numerical optimization of $\ell_W(\t \| \Y)$.  A common technique to overcome these issues is to group the periodogram frequencies into consecutive bins~\citep[e.g.,][Section 10.4]{daniell46,brockwell.davis13}.  That is, assume that \mbox{$K = B \cdot N_B$} is a multiple of the bin size $B$, and consider the average periodogram value in bin $m$,
\begin{equation}\label{eq:psdbin}
\bar Y_m = \frac 1 B \sum_{k \in I_m} Y_k, \qquad I_m = \{k: (m-1)B < k \le mB\}.
\end{equation}
It then follows from Proposition~\ref{prop:expo} that if $\S_i(\f)$ is relatively constant within bins, the distribution of $\bar \Y = (\rv {\bar Y} {N_B})$ can be well approximated by
\begin{equation}\label{eq:bingamma}
\bar Y_m \ind \bar \S_m(\t) \times \textnormal{Gamma}(B, B),
\end{equation}
where $\bar \S_m(\t) = \fs \cdot \S(\bar \f_m, \t)$, $\bar \f_m = \tfrac 1 B \sum_{k \in I_m} \f_k$, and $\textnormal{Gamma}(B, B)$ is a Gamma distribution with mean 1 and variance $1/B$.  This leads to the non-linear least-squares (\NLS) estimator
\begin{equation}\label{eq:nlsq}
\hat \t_{\NLS} = \argmin_\t \sum_{m=1}^{N_B} \big(\bar Y_m - \bar \S_m(\t)\big)^2,
\end{equation}
which is a consistent estimator of $\t$~\citep{norrelykke.flyvbjerg10}. The sum-of-squares criterion~\eqref{eq:nlsq} can be minimized using specialized algorithms such as Levenberg-Marquardt~\citep{levenberg44,marquardt63}, rendering the calculation of $\hat \t_\NLS$ considerably simpler than that of $\hat \t_W$.  However, this gain often incurs a significant loss in statistical precision.

\section{Robust and Efficient Parametric PSD Inference}\label{sec:est}

The choice between \NLS and \MLE estimators imposes a trade-off between computational and statistical efficiency.  In addition, both estimators are highly sensitive to periodic noise which commonly plagues AFM spectral data (Section~\ref{sec:denoise}).  Here we describe a two-stage parametric spectral estimator designed to overcome these issues.

\subsection{Variance Stabilizing Transformation}\label{sec:vs}

To see why the \NLS estimator is sub-optimally efficient, note that the approximate Gamma distribution of the binned periodogram~\eqref{eq:bingamma} can itself be approximated by a Normal with matching mean and variance:
\begin{equation}\label{eq:binnorm}
\bar Y_m \ind \N\big(\bar \S_m(\t), \tfrac 1 B \bar \S_m(\t)^2\big).
\end{equation}
Substituting any constant for the parameter-dependent variance in~\eqref{eq:binnorm} then gives rise to $\hat \t_\NLS$ in~\eqref{eq:nlsq}.  However, by a straightforward application of the statistical delta method~\citep[also known as \emph{propagation of errors}, e.g.,][]{bevington.robinson03}, we note that taking the logarithm of the binned periodogram is a variance-stabilizing transformation:
\[
\var(\log \bar Y_m) \approx \left(\left.\fder {y} {\log y}\right\vert_{y = E[\bar Y_m]}\right)^2 \times \var(\bar Y_m) = \frac{1}{\bar \S_m(\t)^2} \times \frac{\bar \S_m(\t)^2}{B} = 1/B,
\]
such that
\begin{equation}\label{eq:llvs}
Z_m = \log(\bar Y_m) \ind \N\big(\log \bar \S_m(\t), B^{-1}\big).
\end{equation}
Maximizing the likelihood resulting from~\eqref{eq:llvs} leads to the log-periodogram (\LP) estimator
\begin{equation}\label{eq:logp}
\hat \t_{\LP} = \argmin_\t \sumi[m] 1 {N_B} \big(Z_m - \log \bar \S_m(\t)\big)^2.
\end{equation}
This simple sum-of-squares can be effectively minimized by the methods of Section~\ref{sec:bin}, yet with $\hat \t_\LP$ achieving nearly the same precision as $\hat \t_W$.  The \LP estimator is commonly used in statistics to estimate long-range dependence~\citep{geweke.porter-hudak83,robinson95}.  Its asymptotic properties have been derived by~\cite{fay.etal02} and compared favorably therein to the efficient  estimators of~\cite{taniguchi87}.

\begin{remark}
A different variance-stabilizing transformation of the periodogram $\Y$ is to take logarithms before binning, i.e., let $\tilde Z_m = \frac 1 B \sum_{k \in I_m} \log(Y_k)$.  Since the log-Exponential distribution is more Normal than the Exponential itself, a smaller $B$ is required for $\tilde Z_m$ than $Z_m$ for within-bin normality to hold.  However, assuming that approximately $Y_k \iid \bar \S_m(\t) \times \textnormal{Expo}(1)$ for $k \in I_m$, it can be shown that $\var(\tilde Z_m) = \pi^2/6 \times \frac 1 B > \var(Z_m)$.  Therefore, the analogous estimator to~\eqref{eq:logp} with $\tilde Z_m$ in place of $Z_m$ is expected to be less efficient, and indeed this was the case in our numerical experiments.
\end{remark}

\subsection{Periodic Noise Removal}\label{sec:denoise}

The PSD as defined in~\eqref{eq:psd} tacitly assumes that the data are ``purely stochastic''~\citep[in the sense of their Wold decomposition, e.g.,][Theorem 5.1.1]{lindquist.picci15}.  However, the periodogram in Figure~\ref{real_data_psd} has several vertical lines in the \SIrange{e5}{e6}{\hertz} range, suggesting the presence of periodic terms which cannot be explained by a PSD alone.  Indeed, the AFM is a complex instrument operated by extensive electronics, which inevitably leads to
periodic noise from various electrical components and power sources.  Careful engineering can significantly reduce the effects of electronic noise on the final cantilever displacement readings.  However, the residual periodic components shown in Figure~\ref{real_data_psd} can gravely impact PSD parameter estimates as will be demonstrated shortly.  Fortunately, the more severe electronic noise can be easily and automatically removed from the PSD by the following method due to~\cite{fisher29}.

Suppose that the periodogram data $\Y = (\rv Y K)$ contain no periodic components.  Under this null hypothesis, we have
\[
H_0: W_k = \frac{Y_k}{\S_k(\t_0)} \iid \textnormal{Expo}(1),
\]
where $\t_0$ is the true parameter value.  Now consider the maximum jump of the normalized cumulative periodogram density, also known as Fisher's \emph{$g$-statistic}:
\[
\M = \max_{1\le k \le K}\frac{W_k}{\sumi[j] 1 K W_j}.
\]
Under $H_0$, $\M$ is distributed as the maximum distance between the order statistics of $K$ iid uniform random variables, of which the distribution is given by~\citep[e.g.,][Corollary 10.2.2]{brockwell.davis13}
\begin{equation}\label{eq:fisherg}
\Pr(\M > a \| H_0) = \sumi[k] 1 K (-1)^{k+1} {K \choose k}(1-k\cdot a)_+^{K-1}, \qquad x_+ := \max(x, 0).
\end{equation}

\subsection{Proposed Estimator}\label{sec:propest}

The developments above motivate a two-stage parametric spectral density estimator consisting of the following steps:
\begin{enumerate}
	\item \textbf{Preliminary Estimation.} Calculate a preliminary estimate $\hat \t_{\LP}\spp 1$ using the log-periodogram likelihood function~\eqref{eq:logp}.
	\item \textbf{Periodicity Removal.} Calculate $\M$ upon substituting $\hat \t_{\LP}\spp 1$ for the unknown value of $\t_0$, and the $p$-value against large $\M$ using~\eqref{eq:fisherg}.  If the $p$-value is small -- say less than 1\% -- replace the corresponding periodogram ordinate $Y_k$ by a random draw from $\S_k(\hat \t_{\LP}\spp 1) \times \textnormal{Expo}(1)$.  Repeat this procedure until Fisher's $g$-test does not reject $H_0$.
	\item \textbf{Final Estimation.}  Calculate $\hat \t_{\LP}$ on the periodogram obtained from Step 2, from which the unwanted periodicities have been removed.
\end{enumerate}

\begin{remark}
We have opted in Step 2 to replace the periodic outliers by random draws, instead of simply deleting them and repeating Fisher's $g$-test with $K-1$ variables.  This is because the largest of these $K-1$ variables is in fact the second largest of the original $K$, for which~\eqref{eq:fisherg} does not give the right distribution under $H_0$.
\end{remark}

\section{Simulation Study}\label{sec:sim}

\begin{figure}[!htb]
  \centering
  \includegraphics[scale=0.7]{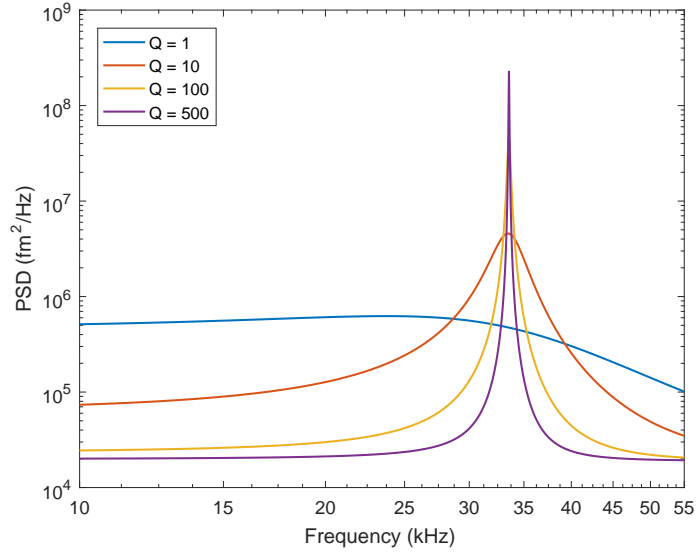}
\caption{Baseline PSDs for various quality factors, over the frequency range used for parameter estimation. Larger quality factors have higher curvature near the resonance frequency.}\label{fig:sim_psd}
\end{figure}

In order to evaluate the parametric spectral estimator proposed in Section \ref{sec:propest}, we consider the following simulation study reflecting a broad range of AFM calibration scenarios. Each simulation run consisted of a \SI{5}{\second} time series sampled at \SI{10}{\MHz} ($N = \num{5e6}$ data points) from the SHO model~\eqref{eq:shopsd} with added white noise,
\begin{equation}\label{eq:show}
\S(\f \| \t) = \Aw + \frac{\kbt/(k\cdot \pi\f_0Q)}{\big[(\f/\f_0)^2-1\big]^2 + \big[\f/(\f_0Q)\big]^2},
\end{equation}
where $\t = (k, \f_0, Q, \Aw)$.  Data was generated using a standard FFT-based algorithm~\citep[e.g.,][]{labuda.etal12b}.  For all simulations, the baseline parameters are displayed in Table~\ref{tab:base}.
\begin{table}[!htb]
\begin{center}
\caption{SHO parameters in baseline environment.} \label{tab:base}
\begin{tabular}{r | l}
  SHO Parameter & Value \\
  \hline
  Temperature & $T = \SI{298}{\kelvin}$ \\
  Stiffness & $k = \SI{0.172}{\newton\per\meter}$ \\
  Resonance Frequency & $\f_0 = \SI{33.533}{\kilo\hertz}$ \\
  Quality Factor & $Q \in \{1,10,100,500\}$
\end{tabular}
\end{center}
\end{table}
All parameters being fixed except $Q \in \{1,10,100,500\}$, the corresponding SHO spectra are displayed in Figure~\ref{fig:sim_psd}.
For each of the four baseline settings, $M = 1000$ datasets were generated, and for each dataset we calculated the three estimators $\hat \t_W$, $\hat \t_\NLS$, and $\hat \t_\LP$.  This was done using only periodogram frequencies in the range $\f_0 \pm \f_0/\sqrt 2$, a range typically provided by the cantilever manufacturer, and outside of which the remaining frequencies provide little additional information about $\t$.  For the \NLS and \LP estimators, the bin size was set to $B = 100$.  For all estimators, the optimization was reduced from four to three parameters by the method of profile likelihood described in Appendix~\ref{app:profl}.

\subsection{Baseline Environment}\label{sec:baseline}

\begin{figure}[!htb]
  \centering
  \includegraphics[width=\textwidth]{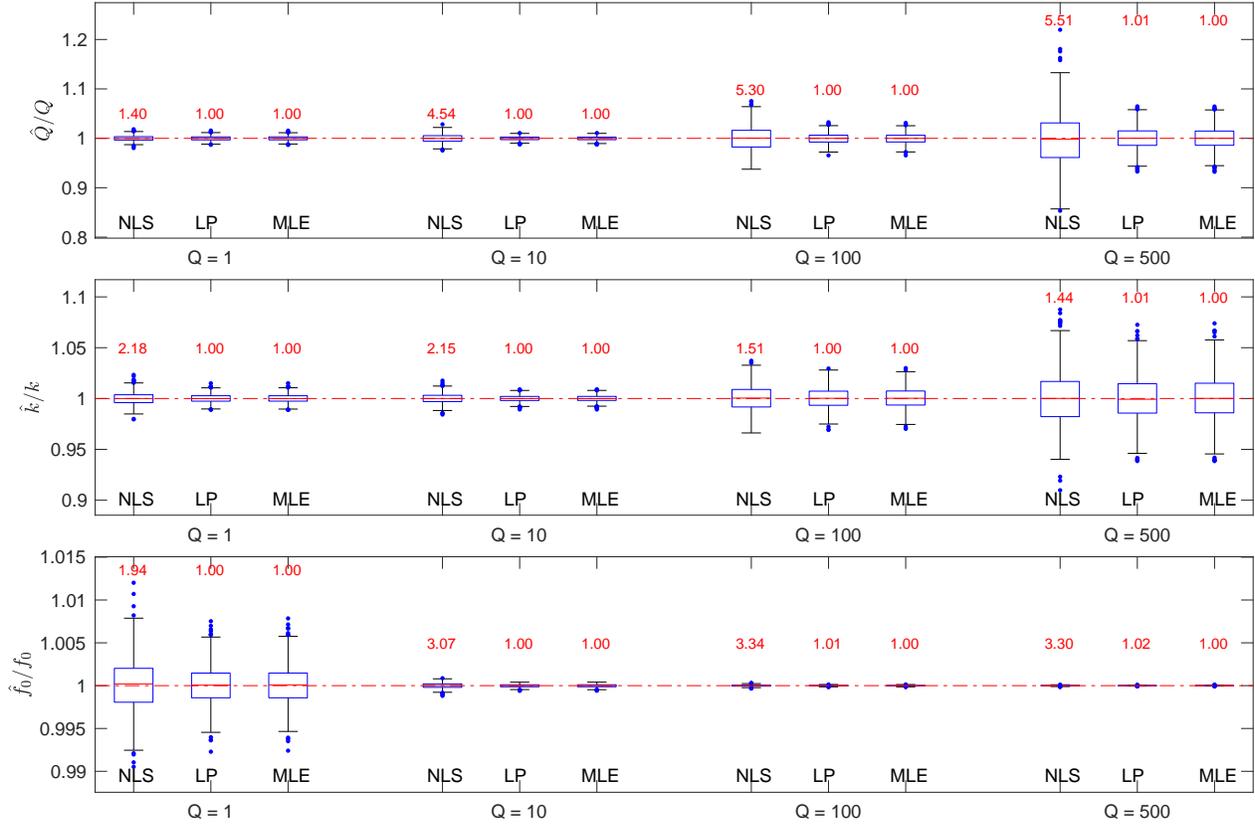}
\caption{Comparison of \NLS, \LP, and \MLE estimators in the baseline simulation environment.  Numbers indicate MSE ratios of the corresponding estimators relative to \MLE.}\label{fig:mse_boxplot}
\end{figure}

Figure \ref{fig:mse_boxplot} displays boxplots for each estimator of each parameter estimate relative to its true value.  The numbers on top of each boxplot correspond to the mean squared error (MSE) ratios between each estimator and the \MLE $\hat \t_W$.  That is, for each of the SHO parameters $\varphi \in (k, \f_0, Q)$ and estimator $j \in \{\NLS, \LP, \MLE\}$, the corresponding MSE ratio in Figure \ref{fig:mse_boxplot} is calculated as
\[
\mathcal R_j(\varphi) = \frac{\sumi[i] 1 M (\hat \varphi_{j}\spp i - \varphi_{0})^2}{\sumi[i] 1 M (\hat \varphi_{W}\spp i - \varphi_{0})^2},
\]
where $\varphi_0$ is the true parameter value and $\hat \varphi_j\spp i$ is its estimate by method $j$ for dataset $i$.

For low quality factor $Q = 1$, the \NLS method has roughly 1.5-2 times higher MSE than the \MLE. For higher values of $Q$, the MSE of \NLS increases to roughly 3-5 times that of \MLE. In contrast, the \LP estimator achieves virtually the same MSE as the \MLE at a small fraction of the computational cost.

\subsection{Electronic Noise Contamination}

In order to assess the impact of electronic noise, a random sine wave of the form \mbox{$D \cdot \sin(2\pi\zeta t + \phi)$} was added to each of the baseline datasets from the simulations above.  The parameters of each sine wave were chosen to mimic the electronic noise in the real AFM data of Figure~\ref{real_data_psd}, a particularly difficult scenario for SHO parameter estimation due to the proximity of the electronic noise to the resonance frequency $\f_0$.  Specifically, the frequency $\zeta$ of each sine wave was generated from a Normal with mean $\f_0 = \SI{33.533}{\kilo\hertz}$ and standard deviation \SI{10}{\hertz}, the phase $\phi$ was drawn uniformly between 0 and $2\pi$, and the amplitude $D$ was set to achieve an approximately ten-fold increase from the maximum value of the baseline PSD near $\f_0$.  The small jitter in the sine wave parameters was added both to mimic the small variations measured in real AFM data, and to investigate the impact of spectral leakage.

\begin{figure}[!htb]
  \centering
  \includegraphics[scale=0.7]{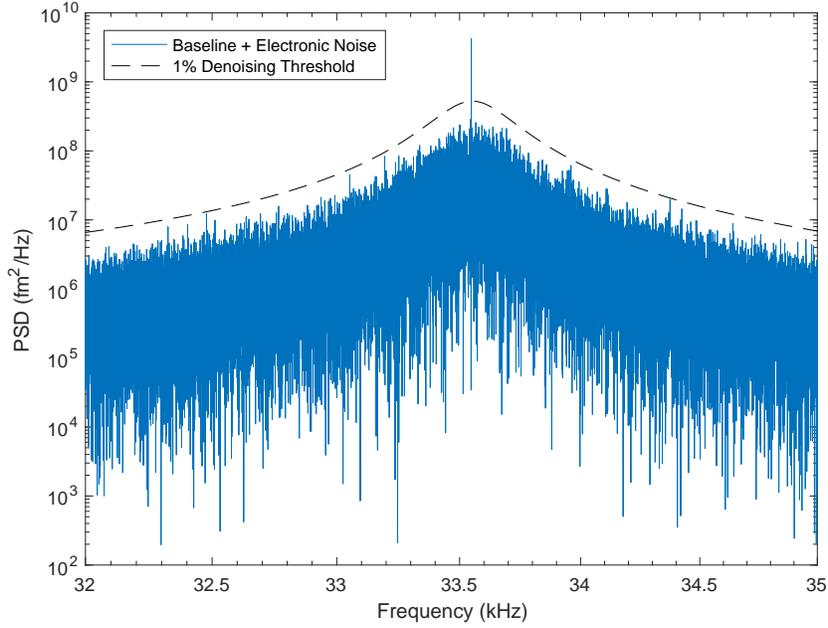}
  \caption{Simulated AFM periodogram with $Q = 100$ and added electronic noise, along with the 1\% denoising threshold prescribed by Fisher's $g$-test. Periodogram ordinates above the threshold are flagged as electronic noise.}
  \label{sine_plot}
\end{figure}

Figure~\ref{sine_plot} displays a simulated dataset with electronic noise contamination.  Also displayed is the 1\% threshold for periodic noise detection by Fisher's $g$-test (Section~\ref{sec:denoise}).  This is calculated by solving numerically for $\Pr(\M > a_\textnormal{cut} \| H_0) = .01$ using~\eqref{eq:fisherg}, then setting the threshold for frequency $\f_k$ to $a_\textnormal{cut} \times \S_k(\hat \t_\LP \spp 1) \cdot \sum_{j=1}^K (Y_j/\S_j(\hat \t_\LP \spp 1))$.  The threshold in Figure~\ref{sine_plot} indicates that most electronic noise detectable to the naked eye can be easily removed by the denoising procedure.

\begin{figure}[!htb]
  \centering
  \includegraphics[width=\textwidth]{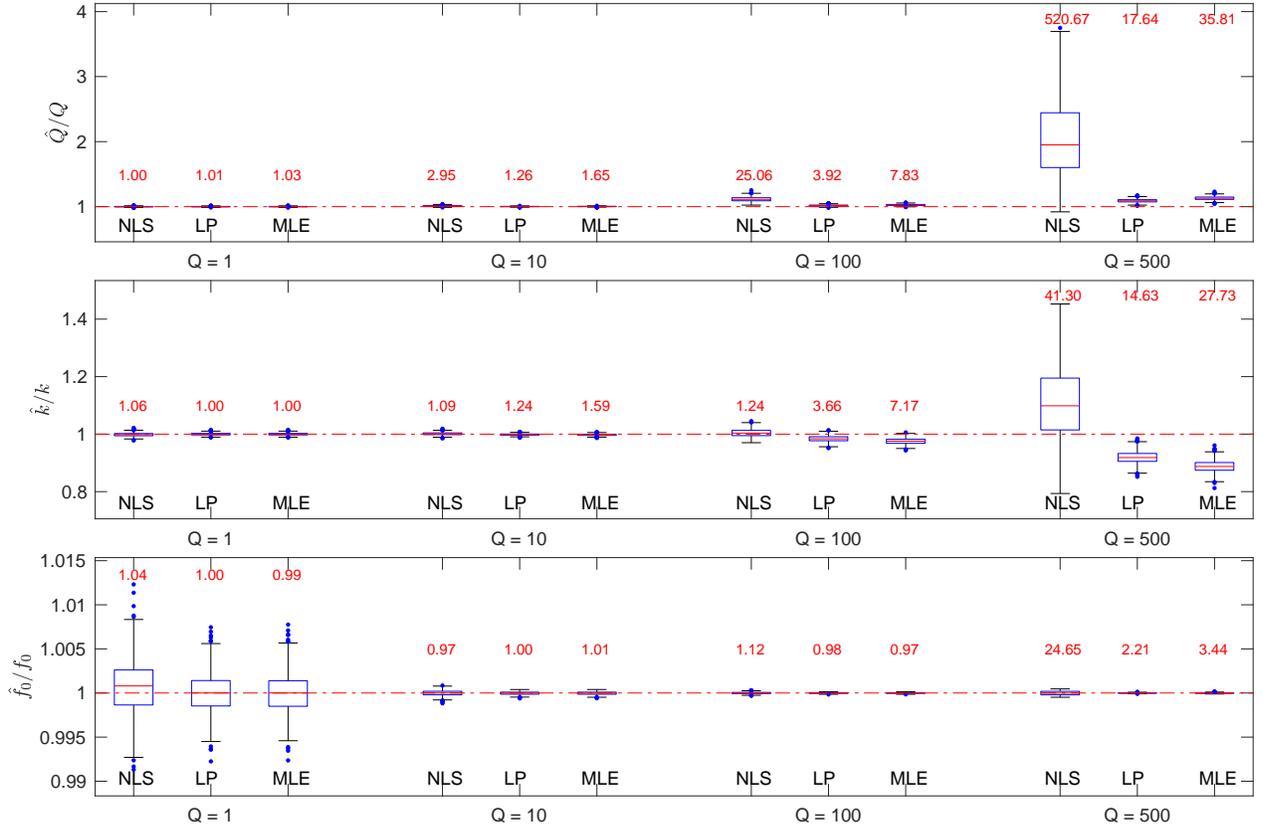}
  \caption{Comparison of \NLS, \LP, and \MLE preliminary estimators (i.e., prior to noise removal), in the noise-contaminated environment.  Numbers indicate MSE ratios of the corresponding estimators relative to their own performance at baseline.}
  \label{fig:mse_boxplot_noise}
\end{figure}

Figure \ref{fig:mse_boxplot_noise} displays boxplots of each parameter estimate relative to its true value in the presence of electronic noise.  To assess the impact of the noise corruption, these estimates do not include the denoising step of Section~\ref{sec:propest}. The numbers in the plot correspond to MSE ratios between the estimator with noise corruption, relative to its own performance in the baseline dataset.  The ratios are thus calculated as
\[
\mathcal R_j(\varphi) = \frac{\sumi[i] 1 M (\hat \varphi_{j,\noise}\spp i - \varphi_{0})^2}{\sumi[i] 1 M (\hat \varphi_{j,\baseline}\spp i - \varphi_{0})^2},
\]
where $\hat \varphi_{j,\baseline}\spp i$ and $\hat \varphi_{j,\noise} \spp i$ are parameter estimates with method  $j$ for dataset $i$ under baseline and noise-contaminated settings, respectively.
At low $Q$, the MSE ratios are close to one, indicating that the estimators are relatively insensitive to the electronic noise.  However, for high $Q$ the effect of the noise is considerably more detrimental, particularly for \NLS.  In all cases, the performance of the \LP estimator is affected the least, indicating it is naturally more robust than \NLS and \MLE to periodic noise contamination, even before the denoising technique is applied.

\begin{figure}[!htb]
  \centering
  \includegraphics[width=\textwidth]{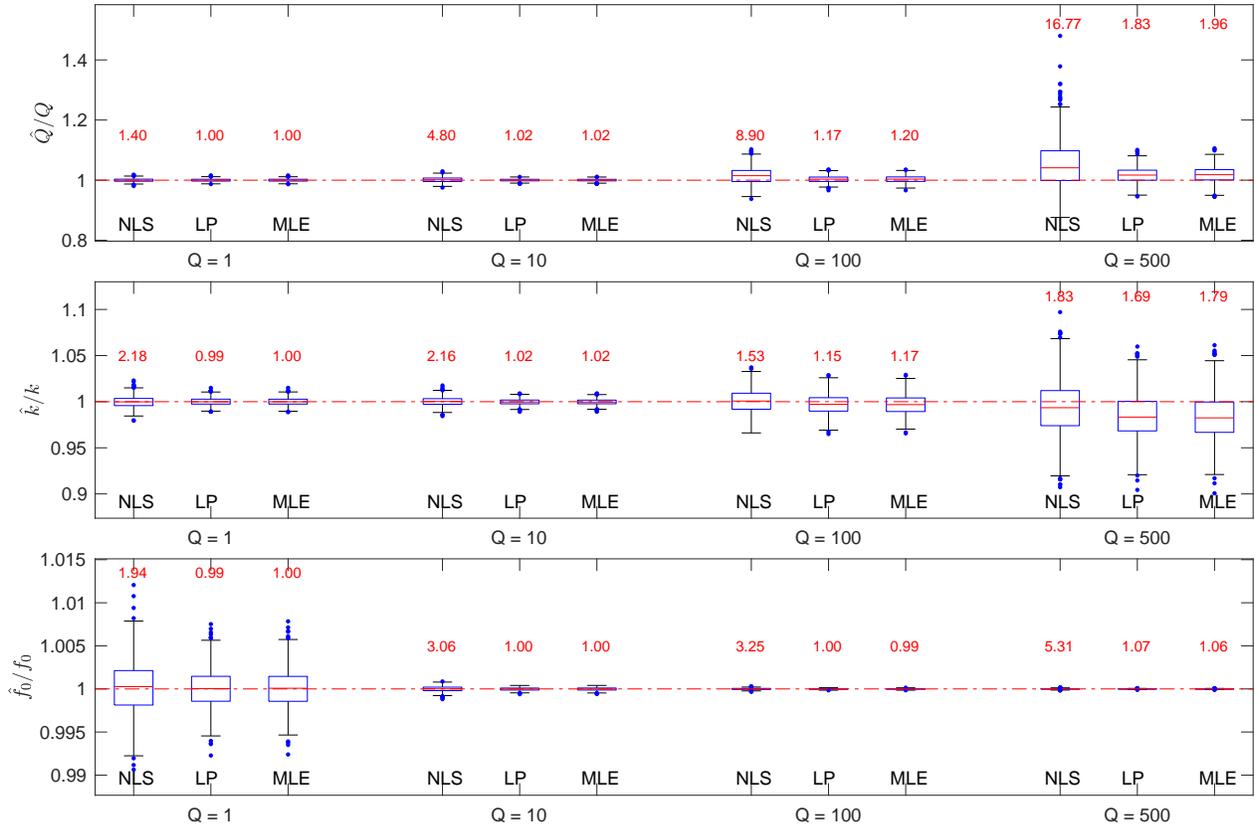}
\caption{Comparison of two-stage \NLS, \LP, and \MLE estimators in the noise-contaminated environment.  Numbers indicate MSE ratios of the corresponding estimators relative to the \MLE at baseline.}\label{fig:mse_boxplot_denoise}
\end{figure}

Figure \ref{fig:mse_boxplot_denoise} displays boxplots for the second-stage parameter estimates, after electronic noise removal.  Each estimator (\NLS, \LP, and \MLE) used its own preliminary fit to determine the noise cutoff value.  Here, the MSE ratios are calculated relative to an ``ideal'' estimator: the \MLE with perfect denoising.  That is, the MSE ratios are
\[
\mathcal R_j(\varphi) = \frac{\sumi[i] 1 M (\hat \varphi_{j,\denoise}\spp i - \varphi_{0})^2}{\sumi[i] 1 M (\hat \varphi_{W,\baseline}\spp i - \varphi_{0})^2},
\]
where $\hat \varphi_{j,\denoise}\spp i$ is the noise-corrected estimate of method $j$ for dataset $i$.

\begin{figure}[H]
  \centering
  {\includegraphics[width=\textwidth]{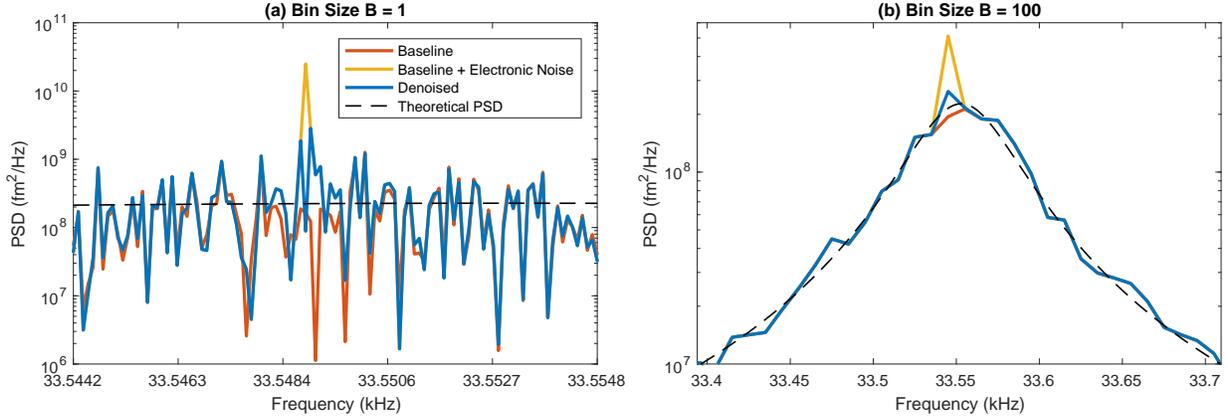}
  \phantomsubcaption\label{biasplot_zoom}\phantomsubcaption\label{biasplot_bin}}
\caption{Magnified view of electronic noise removal process for very high quality factor $Q = 500$.  (a) Electronic noise pushes baseline upwards at several frequencies, due to spectral leakage.  (b) High curvature of PSD near $\f_0$ exposes $\hat Q_{\denoise}$ to upward bias in periodogram bin.}
\label{biasplot}
\end{figure}

In general, the denoising procedure is extremely effective for both \LP and \MLE, but somewhat less so for \NLS (for example, at $Q = 100$ the MSE relative to $\hat Q_{W,\baseline}$ for $\hat Q_{\NLS,\baseline}$ is 5.30, whereas for $\hat Q_{\NLS,\denoise}$ it is 8.90). However, for very high $Q = 500$, the denoising procedure for \LP and \MLE fails to completely remove the upward bias in $\hat Q$ and the downward bias in $\hat k$.  Upon further investigation, Figure~\ref{biasplot} reveals that this is due to spectral leakage. Indeed, a close look at the 50 frequencies on either side of $\f_0$ (Figure~\ref{biasplot_zoom}) shows that several periodogram variables adjacent to the electronic noise at \SI{33.5490}{\kilo\hertz} have been pushed upward by its presence.  The denoising procedure is able to remove the noise at \SI{33.5490}{\kilo\hertz}, but not in the neighboring frequencies.  The net effect after noise correction is a slight upward bias in the binned periodogram (Figure~\ref{biasplot_bin}), which, due to the high curvature of the SHO at $Q = 500$, causes an upward bias in $\hat Q_{\denoise}$.  However, the overall amplitude of the SHO remains unaffected, and since by~\eqref{eq:show} this amplitude is proportional to $\kbt/(k\cdot \pi \f_0 Q)$, the upward bias in $\hat Q_{\denoise}$ is accompanied by a downward bias in $\hat k_{\denoise}$.

\section{Application to Experimental AFM Data}\label{sec:data}

We now turn to the problem of calibrating the AFM cantilever for which the periodogram is displayed in Figure~\ref{real_data_psd}.  The data consist of \SI{5}{\second} of an AC160 Olympus cantilever recorded at \SI{5}{\mega\hertz} ($N = \num{25e6}$ observations).  The objective is to determine the parameters of the best-fitting SHO model to the first cantilever eigenmode (Figure~\ref{fig:psdeig1}).

Calibration of a real AFM cantilever is subject to at least two complications not addressed in the simulations of Section~\ref{sec:sim}:
\begin{enumerate}
\item While the PSDs used in simulation are dominated at low frequencies by white noise, those measured in the real data of Figure~\ref{fig:psdfull} exhibit power-law behavior, $\S(f) \sim 1/\f^\a$ as $\f \to 0$.  This is referred to as ``$1/f$ noise''; it features prominently in AFM experiments~\citep[e.g.,][]{harkey.kenny00,giessibl03,heerema.etal15}, and is due in this case to slow fluctuations of the measurement sensor.  Depending on the exponent, $1/f$ noise induces long-range dependence in the cantilever displacement ($0 < \a < 1$), or even lack of stationarity ($\a \ge 1$).  Failing to account for it can significantly bias SHO parameter estimates.  Fortunately, $1/f$ noise can be dealt with readily by adding a correction term to the SHO model, which becomes
  \[
    \S(\f\|k,\f_0,Q,\Aw,\Af,\a) = \Aw + \frac{\Af}{\f^\a} + \frac{\kbt/(k\cdot \pi\f_0Q)}{\big[(\f/\f_0)^2-1\big]^2 + \big[\f/(\f_0Q)\big]^2}
    .
  \]
  While important at low frequencies, the $1/f$ noise around the first eigenmode (Figure~\ref{fig:psdeig1}) is nearly imperceptible. Consequently, we estimated the first eigenmode's SHO parameters using the simpler model~\eqref{eq:show}. We have constructed a simulation in which $1/f$ noise severely affects SHO parameter estimation in Appendix~\ref{app:showf}.  Relative performance of \LP to \NLS and \MLE estimators was similar to Section~\ref{sec:sim}.

\item In addition to the first  eigenmode at roughly \SI{313}{\kilo\hertz}, the data contain higher eigenmodes corresponding to flexural oscillations of the clamped cantilever beam~\citep{sader98}.  The first of these higher eigenmodes is displayed in Figure~\ref{fig:psdeig2}.  Calibration of higher eigenmodes is of essential importance for popular bimodal and multifrequency AFM imaging techniques~\citep[e.g.,][]{martinez.etal08, garcia.proksch13,herruzo.etal14,labuda.etal16b}, on which we elaborate in the Discussion (Section~\ref{sec:disc}).
\end{enumerate}

\begin{figure}[H]
  \centering
  \includegraphics[width=.6\textwidth]{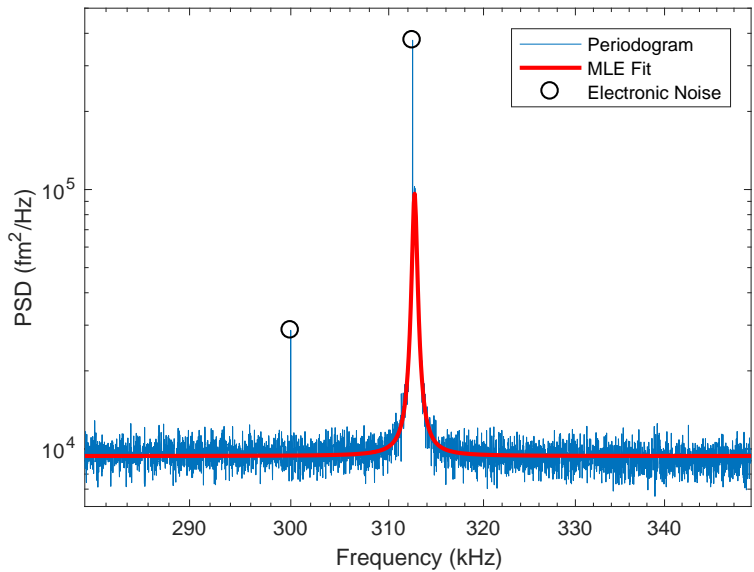}
\caption{Periodogram (averaged by bins of size $B =100$) and \MLE fit (\NLS and \LP fits are virtually indistinguishable).  The circles indicate frequencies flagged as electronic noise.}
\label{real_fit}
\end{figure}

Figure~\ref{real_fit} displays the periodogram of the AFM data from Figure~\ref{real_data_psd} over the frequency range used for parameter estimation.  The electronic noise at \SI{312.5}{\kilo\hertz} and \SI{300.0}{\kilo\hertz} was easily removed with Fisher's $g$-statistic. Table~\ref{real_table} displays parameter estimates and standard errors for \NLS, \LP, and \MLE methods, the first two being calculated with bin size $B = 100$. For \LP and \MLE, standard errors are calculated by inverting the observed Fisher information matrices corresponding to~\eqref{eq:logp} and~\eqref{eq:whittle}.  For \NLS, standard errors are obtained by the sandwich method~\citep[e.g.,][]{freedman06}.

For this particular dataset, the \NLS, \LP, and \MLE estimators are fairly similar, all being within one standard error of each other. This is because the difference between the estimators is largely driven by the relative amplitude of the SHO peak to its base.  Here this ratio is about 10, which is similar to the $Q = 10$ scenario examined in Section~\ref{sec:sim}.  Indeed, repeating the simulations of Section~\ref{sec:sim} with true parameters values taken as the \MLE estimates in Table~\ref{real_table} produced similar results to the aformentioned scenario, i.e., indistinguishable \LP and \MLE estimators having three times smaller MSE than \NLS.
\begin{table}[!htb]
\begin{center}
\caption{Real AFM cantilever parameter estimates and standard errors.} \label{real_table}
\begin{tabular}{ c |  c  c  c }
  & $\hat f_0$ (\si{\kilo\hertz}) & $\hat Q$ (unitless) & $\hat k$ (\si[per-mode=symbol]{\newton\per\meter}) \\
\hline
\NLS & 312.703 (.0043) & 603.01 (14.28) & 57.52 (0.91) \\
\LP & 312.701 (.0047) & 595.40 (12.74) & 57.16 (0.81) \\
\MLE & 312.700 (.0048) & 593.58 (12.66) &  57.20 (0.81)
\end{tabular}
\end{center}
\end{table}

\subsection{Bin Size}

While for this dataset there is little difference between the various estimators, \NLS and \LP can be substantially faster than \MLE due to periodogram frequency binning.  In practice, the choice of bin size affects both computational efficiency and approximation accuracy.  Large bin sizes can group periodogram variables with very different amplitudes, thus invalidating the Gamma approximation to $\bar Y_m$ in~\eqref{eq:bingamma}.  On the other hand, small bin sizes can strain the Normal approximations to $\bar Y_m$ and $\log(\bar Y_m)$ in~\eqref{eq:binnorm} and~\eqref{eq:llvs}.

\begin{figure}[H]
  \centering
  \includegraphics[width=\textwidth, trim={5mm 0mm 0mm 0mm},clip]{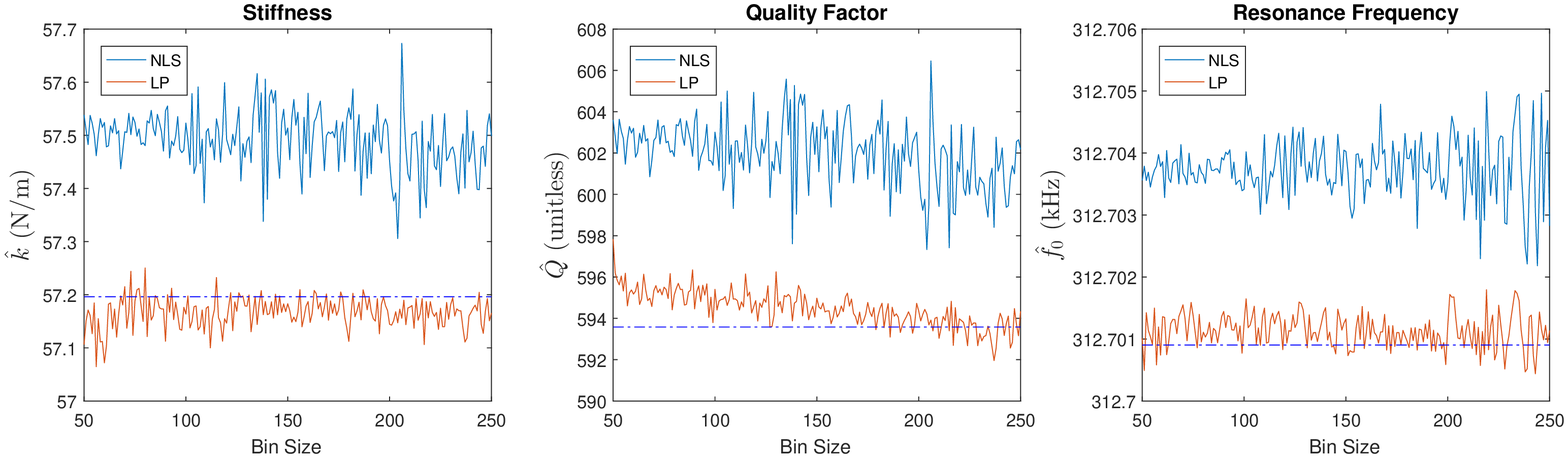}
\caption{\NLS and \LP parameter estimates for real AFM cantilever, for bin size $B = \textrm{\numrange{50}{250}}$.  The \MLE estimate (dashed line) is added for reference.}\label{bin_plot}
\end{figure}

To investigate the impact of bin size, Figure~\ref{bin_plot} plots \NLS and \LP estimators for the values of $B = \textrm{\numrange{50}{250}}$. The behavior of \NLS is considerably more erratic, presumably due to small changes in the bin end points having larger impact on $\bar \S_m(\t)$ than $\log \bar \S_m(\t)$.  Note that the downward trend in $\hat Q$ is caused by increased flattening of the periodogram curvature as bin size increases.

\section{Discussion} \label{sec:disc}

Parametric spectral density estimation plays a key role in AFM cantilever calibration.  We have proposed a two-stage parameteric spectral estimator having statistical efficiency comparable to \MLE at a small fraction of the computational cost, robust to most adverse effects of periodic noise contamination (except perhaps for very sharply peaked spectra).  As spectral leakage due to binning affects the choice of bin size, a possible direction for future work is the construction of variable bin sizes, to be determined after the preliminary fit.
Another line of future investigation is the calibration of higher eigenmodes.  In principle, this can be done by fitting separate SHO models to each successive eigenmode.  However, as the peak amplitude of these higher modes gets closer and closer to the noise floor, the accuracy of separate SHO estimators rapidly deteriorates.  Instead one might wish to combine SHO models on the basis of hydrodynamic principles~\citep{vaneysden.sader06,clark.etal10} and other scaling laws~\citep{labuda.etal16a}.

\appendix

\section{Profile Likelihood for SHO Fitting}\label{app:profl}

We begin by reparametrizing the SHO model~\eqref{eq:show} as
\begin{align*}
  \S(\f \| \t) & = \Aw + \frac{\kbt/(k\cdot \pi\f_0Q)}{\big[(\f/\f_0)^2-1\big]^2 + \big[\f/(\f_0Q)\big]^2} \\
               & = \tau \times \left\{\Rw + \frac{1}{\big[(\f/\f_0)^2-1\big]^2 + (\f/\g)^2}\right\} = \tau \cdot \G(\f \| \ee),
\end{align*}
where $\tau = \kbt/(k\cdot \pi\f_0Q)$, $\Rw = \Aw/\tau$, $\g = \f_0Q$, and $\ee = (\f_0, \g, \Rw)$.  The objective function for the \NLS estimator then becomes
\[
  \obj_\NLS(\tau, \ee) = \sum_{m=1}^{N_B}\big(\bar Y_m - \tau \cdot \bar\G_m(\ee)\big)^2,
\]
where $\bar \G_m(\ee) = \fs \cdot \G(\bar \f_m, \ee)$.  For any fixed value of $\ee$ the value of $\tau$ which minimizes $\obj_\NLS(\tau, \ee)$ is
\[
  \hat \tau(\ee) = \argmin_\tau \obj_\NLS(\tau, \ee) = \frac{\sum_{m=1}^{N_B} \bar \G_m(\ee) \cdot \bar Y_m}{\sum_{m=1}^{N_B} \bar \G_m(\ee)^2}.
\]
It follows that by setting $\hat \ee_\NLS = \argmin_{\ee} \obj_\NLS(\hat \tau(\ee), \ee)$ and $\hat \tau_\NLS = \hat \tau(\hat \ee_\NLS)$, we have $(\hat \tau_\NLS, \hat \ee_\NLS) = \argmin_{(\tau,\ee)} \obj_\NLS(\tau, \ee)$.  We can then recover the corresponding estimator $\hat \t_\NLS = \argmin_\t \obj_\NLS(\t)$ by applying the inverse transformation $Q = \g/\f_0$, $k = \kbt/(\tau \cdot \pi \g)$, and $\Aw = \Rw \cdot \tau$.  Thus, we have obtained $\hat \t_\NLS$ at the cost of the three parameter optimization of $\obj_\NLS(\hat \tau(\ee), \ee)$, rather than the four parameter direct optimization of $\obj_\NLS(\t)$.

An analogous ``profiling'' procedure can be applied to the \LP and \MLE estimators, in order to reduce the optimization problem from four parameters to three.  For \LP, the objective function is
\[
  \obj_\LP(\t, \ee) = \sum_{m=1}^{N_B} \big(Z_m - \log(\tau) - \log \bar \G_m(\ee)\big)^2,
\]
for which
\[
  \hat \tau(\ee) = \argmin_\tau \obj_\LP(\t, \ee) = \exp \left\{\frac 1 {N_B} \sum_{m=1}^{N_B} \big(Z_m - \log \bar \G_m(\ee)\big)\right\}.
\]
Similarly, the objective function for the \MLE estimator is
\[
  \obj_\MLE(\t, \ee) = \sum_{k=1}^K \left(\frac{Y_k}{\tau \cdot \G_k(\ee)} + \log(\tau) + \log \G_k(\ee)\right),
\]
where $\G_k(\ee) = \G(\f_k \| \ee)$, for which
\[
  \hat \tau(\ee) = \argmin_\tau \obj_\MLE(\t, \ee) = \frac 1 K \sum_{k=1}^K \frac{Y_k}{\G_k(\ee)}.
\]

\section{$\bm {1\,/f}$ Noise} \label{app:showf}

\begin{figure}[!htb]
  \centering
  \includegraphics[width=\textwidth]{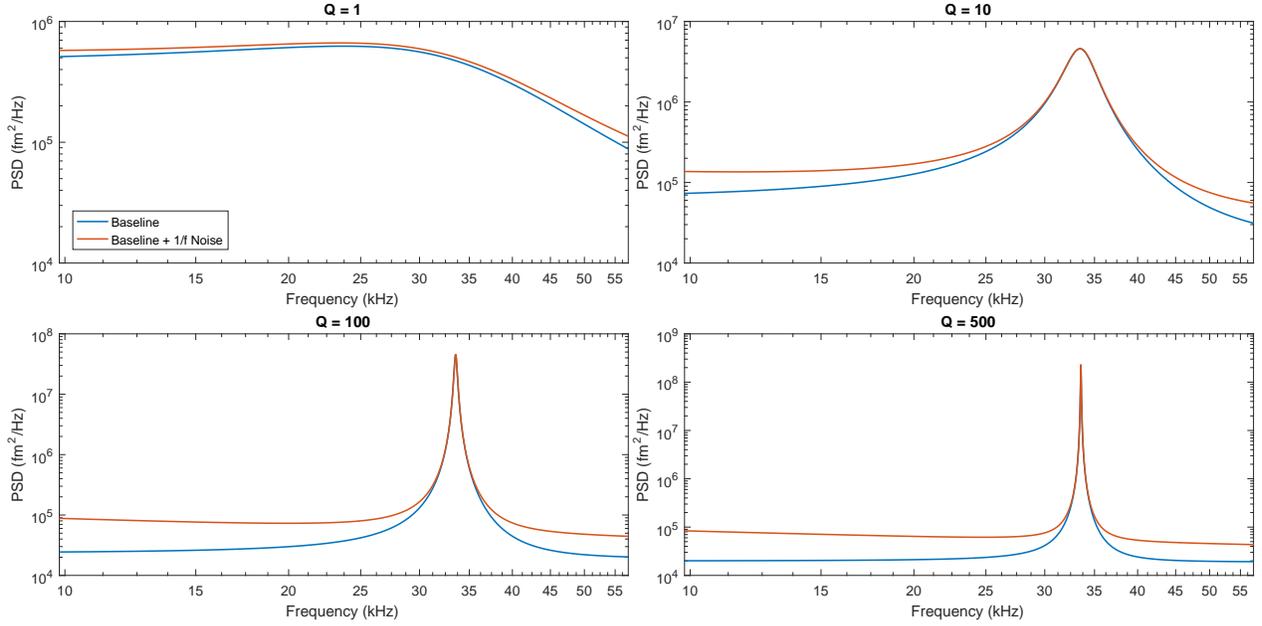}
  \caption{Comparison of $1/f$ noise to baseline power spectral density for various quality factors.}
  \label{pink_distortion_plot}
\end{figure}

The presence of $1/f$ noise is a common feature of AFM power spectra.  This type of noise typically arises from slow fluctuations of the laser and photodiode sensor~\citep{labuda.etal12a} and other long-term cantilever instabilities~\citep{paolino.bellon09}.  It is manifested by a power law behavior at low frequencies, $\S(\f) \sim 1/\f^\a$ as $\f \to 0$.  Thus, a PSD model for the SHO with both white noise and $1/f$ noise contamination is
\begin{equation}\label{eq:showf}
  \S(\f\|k,\f_0,Q,\Aw,\Af,\a) = \Aw + \frac{\Af}{\f^\a} + \frac{\kbt/(k\cdot \pi\f_0Q)}{\big[(\f/\f_0)^2-1\big]^2 + \big[\f/(\f_0Q)\big]^2},
\end{equation}
where $\a$ and $\Af$ are the $1/f$ noise exponent and amplitude parameters, respectively.  While the SHO estimates for the real AFM data in Figure~\ref{real_data_psd} were not impacted by the $1/f$ noise, here we construct a simulation study in which they are.  Namely, we use the baseline parameters described in Section~\ref{sec:sim}, to which we add $1/f$ noise with parameters $\a = 0.55$ and $\Af = \SI{1.0e7}{\square\femto\meter\per\hertz}$.  Baseline and noise contaminated power spectra are displayed in Figure~\ref{pink_distortion_plot}.
To quantify the severity of the $1/f$ noise, Table~\ref{pink_effect} displays the asymptotic relative bias (i.e., as $N \to \infty$) due to fitting the \mbox{SHO + white noise} model~\eqref{eq:show} without accounting for the $1/f$ noise in Figure~\ref{pink_distortion_plot}.  This was calculated by a direct curve-fitting procedure.
\begin{table}[!htb]
\begin{center}
\caption{Asymptotic relative bias for each parameter resulting from failure to account for the $1/f$ noise in Figure~\ref{pink_distortion_plot}.} \label{pink_effect}
\begin{tabular}{ l |  c  c  c }
  & $\f_0$ & $Q$ & $k$ \\
\hline
$Q = 1$ & 1.02 & .92 & .92 \\
$Q = 10$  & 1.00 & .89 & .98 \\
$Q = 100$  & 1.00 & 1.05 &  1.00 \\
$Q = 500$  & 1.00 & 1.10 &  1.00
\end{tabular}
\end{center}
\end{table}
\begin{table}[!htb]
\begin{center}
\caption{Relative MSE of \NLS and \LP estimators to \MLE in the presence of $1/f$ noise. For $Q = 1$ some of the estimators failed to converge, such that ratios are based on 1000 datasets for \NLS, 996 for \LP, and 768 for \MLE.} \label{pink_table}
\begin{tabular}{  l  |  c  c  c  c  }
 & Method & $f_0$ MSE Ratio & $Q$ MSE Ratio & $k$ MSE Ratio \\
\hline
Q = 1 & \NLS & 2.23 & 1.97 & 2.42  \\
& \LP & 1.16 & 1.51 & 1.78 \\
& & & & \\
Q = 10 & \NLS & 2.59 & 3.74 & 1.97 \\
& \LP  & 1.01 & 1.01 & 1.01 \\
& & & & \\
Q = 100 & \NLS & 3.07 & 5.27 & 1.48 \\
& \LP  & 1 & 1 & 1.01 \\
& & & & \\
Q = 500 & \NLS & 3.25 & 5.94 & 1.62 \\
& \LP  & 1.01 & 1 & 1 \\
\end{tabular} \\
\end{center}
\end{table}
While the bias on $\f_0$ and $k$ is relatively small, for $Q$ it is on the order of \numrange{5}{10}\%.

To evaluate the different estimators, $M = 1000$ datasets are generated under each setting as in Section~\ref{sec:sim}, and \NLS, \LP, and \MLE parameter estimates are calculated for each dataset.  For the \NLS and \LP estimators the bin size was $B = 100$.  Table~\ref{pink_table} displays the parameter-wise MSE ratio for \NLS and \LP estimators relative to the \MLE.  For moderate $Q \ge 10$, the performance of the \LP estimator is virtually the same as the \MLE, and \numrange{1.5}{5} times superior than that of \NLS. For very low $Q = 1$, the $1/f$ noise in Figure~\ref{pink_distortion_plot} is almost undetectable, leading to parameter identifiability issues in the fitting algorithms.  In such a setting we recommend to first estimate the $1/f$ parameters separately from the low frequency periodogram values, then estimate the SHO and white noise parameters with $\hat \a$ and $\hat \Af$ fixed.

\section*{Supplementary Materials}
\textbf{Software:} All code for the various PSD fitting algorithms is available at \\ \ifblindversion [\textit{URL withheld in compliance with double-blind policy}]. \else \url{https://github.com/mlysy/realSHO}. \fi

\bibliographystyle{jasaref}
\bibliography{literature_ref2}


\end{document}